\documentclass[a4paper,english]{article}
\usepackage{hyperref}
\usepackage{tahasty}
\usepackage{enumerate}   

\title{Bipartite Entanglement of Deformed Coherent States}

\author[1\footnote{rouabah.taha@umc.edu.dz}]{Mohamed Taha Rouabah}
\author[1]{Noureddine Mebarki}
\affil[1]{Laboratoire de Physique Math\'ematique et Subatomique,
Fr\`eres Mentouri University Constantine-1, 25017 Constantine, Algeria. }

\providecommand{\keywords}[1]
{
  \small	
  \textbf{{Keywords: }} #1
}

\date{}

\begin{document}

\maketitle

\begin{abstract}
A $q$-deformed Weyl-Heisenberg algebra is used to define a deformed displacement operator giving rise to a naturally normalized nonlinear coherent states type as non-classical non-Gaussian quantum states playing a significant role in the entanglement distillation, quantum computation with cluster states and quantum error correction. Robust maximally entangled deformed coherent states are studied and the effect of such a deformation on the amount of the entanglement is discussed. The analogy between environment decoherence and algebra deformation is made through the deformation parameter. 
%
\end{abstract}



\markboth{Mohamed Taha Rouabah, Noureddine Mebarki
}
{Bipartite Entanglement of Deformed Coherent States}

\section{Introduction}
The interaction of quantum systems involved in quantum information processing with the surrounding environment may affect their physical properties, resulting in  decoherence \cite{Karasik2007}. From a mathematical point of view, decoherence effects affecting a quantum system may be represented by a modification of its symmetry. This could be performed through a correction on the mathematical formalism consisting in a small deformation introduced to the commutation relation defining the algebra that describes the system.
\\

The notion of algebra deformation is very familiar to mathematical physicist. Many studies devoted to the investigation of a deformed quantum oscillator algebra appeared as early as seventies, involving deformed creation and annihilation operators known as the $q-$oscillator algebra or the $q-$deformed algebra \cite{Arik1976,Kulish1983,Jimbo1985,Drinfeld1986,Biedenharn1989, Macfarlane1989, Curtright1990}. From a mathematical point of view, the $q-$oscillators lader operators are shown to have a structure of a non-trivial Hopf algebra \cite{Polychronakos1990}.
However, the physical relevance of $q-$deformed creation
and annihilation operators is not always very transparent in the studies that have been published on the subject so far. Therefore it is important to emphasize that there are -- from our point of view -- at least, two main properties which make $q-$oscillators interesting objects for physics. The first is the fact that they constitute a fundamental tools of completely integrable theories \cite{Jimbo1986}. The second, which is one of the points we are interested in, concerns the connection between the $q-$deformation and nonlinearity in the context of coherent states (CS) \cite{Manko1998}.\\

Since 1926, at the beginning of quantum mechanics, Schr\"odinger was interested to the study of quantum states that mimic their classical analogs and defined CS as the states minimizing the Heizenberg uncertainty relation  \cite{Schrodinger1926,Mandel1995}. They were rediscovered by Klauder at the beginning of 1960s \cite{Klauder1963a,Klauder1963b}, then by Sudarshan and Glauber in a series of papers for the description of the coherence phenomenon in lasers \cite{Sudarshan1963,Glauber1963a,Glauber1963b,Glauber1963c}. A decade later, Barut and Girardello have introduced coherent states defined as eigenstates of the  annihilation operator \cite{Barut1971}, then Gilmore and Parelomov have constructed coherent states by the application of a displacement operator on the vacuum state \cite{Perelomov1972}.\\

%
%
The construction of nonlinear coherent states (NCS) as non-classical non-Gaussian quantum states has an important physical motivation and can play a significant role in the entanglement distillation, quantum computation with cluster states , loophole-free Bell tests with the continuous variables and quantum error correction. 
It can also be used to improve the quantum communication protocols in quantum information processing and the optimal estimation of losses. Moreover, some features such as amplitude squeezing and self-splitting which are accompanied by pronounced quantum interference effects contributing to the application in optimizing quantum information in binary (or multibinary) communication are investigated through the construction of NCS \cite{Leon-Montiel2011}. 
It is worth to mention that practically, the generation of the NCS are explored in the study of a single-atom laser where the stationary state of this later is shown to be a phase-averaged NCS \cite{Kilin2012}, as well as in a micromaser under the intensity-dependent Jaynes-Cummings model \cite{Naderi2005}. An exciton in a wide quantum dot  interacting with two laser beams can also be prepared in a NCS and a trapped ion \cite{Harouni2008}. Furthermore, the generation of a non-classical NCS light source is one of the major requirements of quantum information processor associated with the modern semiconductor fabrication technology. In the last few years, it is found that the optomechanical systems can be used  for preparing non-classical states where the generation of the phonons NCS can be achieved via the two-photon or multi-photon blockade observed in cavity quantum electrodynamics (QED) systems with a quantum dot in photonic crystal cavities strongly coupled to a photonic crystal resonator and the dark state of the system which can be realized by tuning the frequency of the weak driving field. In fact,  The optomechanical system can evolve into a dark state due to the damping of the microcavity and the decay of the dressed quantum dot at selected frequencies in the photonic band gap materials. As a result, the phonon mode of the mechanical resonator can be prepared in a NCS \cite{Yan2016}. \\

In this paper, 
we use a weak deformation approximation  to construct a deformed CS displacement operator. In section two, we present our mathematical formalism and highlight some interesting properties of our deformed coherent states especially the analogy with their nonlinear counterpart. In the third section, the bipartite concurrence of the maximally entangled deformed coherent states is discussed. In section four we present the results of our numerical study of the bipartite entangled coherent states concurrence before  drawing our conclusions.

\section{Mathematical formalism}\label{sec:math}
In the Fock  representation, Schr\"odinger-Klauder-Glauber coherent states called {\it standard coherent states} or also {\it canonical coherent states} are given by 
\begin{align}
|\alpha\rangle = e^{-\frac{|\alpha|^2}{2}}\sum_{n=0}^{\infty} \dfrac{\alpha^n}{\sqrt{n!}} |n\rangle \ .
\end{align}
where $\alpha = |\alpha|e^{i\phi}$ is a complex parameter such that $\langle \alpha|\alpha\rangle = 1$ and $|n \rangle$ is the number operator eigenstate. The quantum harmonic oscillator CS  $|\alpha\rangle$ reproduce in averaging the same classical behavior. This unique specificity is due to numerous properties of CS:
\begin{enumerate}[(i)]
\item The CS $|\alpha\rangle$ minimizes the Heisenberg uncertainty relation:
\begin{align}
\langle \Delta Q \rangle_\alpha \langle \Delta P \rangle_\alpha  = \frac{\hbar}{2} \ ,
\end{align}
with $\langle \Delta X \rangle_\alpha = \sqrt{\langle \alpha| X^2|\alpha \rangle - \langle\alpha| X|\alpha \rangle^2} $, ($X= Q, P$) where $Q$ and $P$ are the oscillator position and momentum operators respectively.
\item
 $|\alpha\rangle$ is eigenstate of the annihilation operator $a$ with eigenvalue $\alpha$:
\begin{align}
a |\alpha\rangle = \alpha|\alpha\rangle, \hspace{0.5cm} \alpha \in \mathbb{C} \ ,
\end{align}
where $a = \dfrac{m\omega Q + iP}{\sqrt{2m\hbar\omega}}$ satisfies the commutation relation
\begin{align}\label{eq:WH}
 [a,a^+] = aa^+ - a^+a = 1 \ ,
\end{align} 

\item
The CS $|\alpha\rangle$ is obtained from the fundamental state $|0\rangle$ with a unitary transformation of the Weyl-Heisenberg group $D(\alpha)=e^{\alpha a^+ - \bar\alpha a}$ called displacement operator:
\begin{align}
|\alpha\rangle = e^{\alpha a^+ - \bar\alpha a}|0\rangle \ .
\end{align}
\item
The coherent states $\{|\alpha\rangle\}$ form a complete set in the Hilbert space, such that
\begin{align}
\dfrac{1}{\pi} \int_{\mathbb{C}}  {\rm d \{Re}(\alpha)\}~ {\rm d\{ Im} (\alpha)\} \ |\alpha\rangle\langle \alpha| = I \ ,
\end{align}
where $I$ stands for the identity operator.
\item
The probability $p(n)$ to be in the number operator eigenstate $|n\rangle$ is time independent and has a Poisson distribution
\begin{align}
p(n)= |\langle n|\alpha\rangle|^2 = \dfrac{\mid\alpha\mid^{2n}}{n!} e^{-\mid\alpha\mid^2} \ .
\end{align}
\end{enumerate}
Those properties were taken as a basis for all the generalizations of the concept of standard coherent states.\\

Moreover, the quantum harmonic oscillator model was largely used in quantum mechanics. Indeed, its analytical solution is well known and any quantum system around its equilibrium position can be modeled with a harmonic oscillator. 
However, for more complicated systems, e.g., out of equilibrium systems, the harmonic description is no longer sufficient and additional terms need to be introduced in the position coordinates. An interesting alternative consists on maintaining the same algebraic structure of the harmonic Hamiltonian and introducing a {\it deformed} creation and annihilation operators such that, when rewritten in term of those redefined operators, the new Hamiltonian reproduces the anharmonic behavior of the system.\\ 

In what follows we introduce a type of  deformation where the related creation and annihilation operators denoted by $b$ and $b^+$ respectively define a modified Weyl-Heisenberg algebra satisfying the $q-$deformation commutation relation \cite{Biedenharn1989, Macfarlane1989}
\begin{equation}\label{eq:d-WH1}
[b,b^+]_q = bb^+ - q~b^+b = 1 \ ,
\end{equation}
where the deformation parameter $q$ is taken to be real. In our approach, we use a week deformation approximation where $q= 1 + \varepsilon$ with $\varepsilon \ll 1$, such that we recover the non-deformed case for $\varepsilon \rightarrow 0$. 
In this case one can show that up to $\mathcal{O}(\varepsilon^{2})$ we have:
\begin{align}\label{eq:d-WH2}
[b,b^+]&= bb^+ - b^+b \ , \nonumber\\
			&=  1 + \varepsilon~b^{\dag}b \ .
\end{align}
As a representation of the deformed operators  $b$ and $b^+$ satisfying the commutation relation \eqref{eq:d-WH2} one has :
\begin{align}
b &=   a + \frac{1}{4} \varepsilon a^+ a^2 + \mathcal{O}(\varepsilon^2) \label{eq:d-op1}\ , \\
b^+ &=   a^+ + \frac{1}{4} \varepsilon  a^{\dag2}  a + \mathcal{O}(\varepsilon^2) \label{eq:d-op2}\ .
\end{align}
In this representation the deformed number operator will be defined as:
\begin{equation}
\hat n_d = b^+b = \hat n + \varepsilon (\hat n+ \frac{1}{2}\hat n^2) + \mathcal{O}(\varepsilon^2) \ ,
\end{equation}
where $\hat n =a^+a$ is the non-deformed number operator. \\

It is worth to mention that the so called $f-$deformed oscillators (generalization of $q-$deformed oscillators), were interpreted as nonlinear oscillators  corresponding classically to a frequency dependence of the oscillation amplitude. In the framework of nonlinear coherent states (NCS), Man'ko  defined an $f-$deformed oscillator with the creation and annihilation operators \cite{Manko1993a,Manko1993b,Manko1998}
\begin{align}
b &= a f (\hat n)=f(\hat n+1)a \label{eq:non-lin1}\ ,\\
b^+ &= f (\hat n)a^+=a^+ f (\hat n +1) \label{eq:non-lin2}\ ,
\end{align} 
satisfying :
\begin{align}\label{eq:non-lin3}
[\hat n,b] &=-b, \hspace{1cm} \, \nonumber\\
[\hat n,b^+] &= b^+  \ , \nonumber\\
 [b,b^+] &= (\hat n+1)f^2(\hat n+1) - \hat n f^2(\hat n) \ .
\end{align}
Comparing Eqs.\eqref{eq:d-op1},\eqref{eq:d-op2} and Eqs.\eqref{eq:non-lin1},\eqref{eq:non-lin2}, one can show that in our case the $f$-deformation function has the following expression 
\begin{align}\label{eq:non-lin5}
f(\hat n)= 1 + \frac{1}{4}~\varepsilon~ (\hat n-1),
\end{align}
such as the non-deformed limit is recovered for $f(\varepsilon \rightarrow 0) = 1$.
 Thus, the CS which we will construct starting from our deformed algebra could be interpreted as NCS. We remind the reader the major importance of NCS in the study of nonlinear potentials systems where their non-classical state description  of the electromagnetic field, quantum optics and the atomic center of mass displacement was successful  \cite{Meekhof1996,Monroe1996,Manko2000,Sivakumar2000,Sunilkumar2000,Kis2001,Freitas2014}.
%
%
In fact, the interest for the generalization of CS pertinent for the nonlinear potentials have started since the beginning of 1970s \cite{Barut1971,Perelomov1972,Roy1982}.
 Nieto and co-workers constructed CS corresponding to the P\"oschl-Teller one dimensional potential as the states minimizing the uncertainty relation between the canonical coordinates $Q$ and $P$ \cite{Nieto1978}. Two decades later Gazeau and Klauder proposed a generalization to one dimensional systems with discrete and continue spectrum \cite{Gazeau1999}. 
 The CS corresponding to the trigonometric and modified P\"oschle-Teller potentials were derived in Ref.\cite{Santos2011} and  those concerning the Morse potential in Ref.\cite{Santos2012}. Man'ko {\it et al} \cite{Manko1997} and  Filho \cite{deMatos1996}, have introduced NCS as eigenstates of the deformed annihilation operator whereas the displacement operator NCS were derived in  Ref.\cite{Roy2000a,Roy2000b}. \\

The derivation of CS from displacement operator was generalized to deformed oscillators at the end of 1990s. In the same spirit as in Ref.\cite{Aniello2000} we introduce a deformed displacement operator $D_d(\alpha)$ such that :
\begin{align}\label{eq:dDOp}
 D_d(\alpha) = \exp\left(\alpha b^+ - \bar{\alpha}b	\right) \ ,
\end{align}
generating deformed coherent states (DCS) $|\alpha\rangle_d $ defined as
\begin{align}\label{eq:dCS1}
 |\alpha\rangle_d = D_d(\alpha)| 0\rangle_d \ ,
\end{align}
where $| 0\rangle_d$ is the deformed vacuum state.
Starting from the fact that $\exp(\bar \alpha b) |0\rangle_d = |0\rangle_d$ and using the BCH formula, straightforward simplifications (see appendix (\ref{apdx:DCS})) lead to the following expression for the DCS defined in Eq.\eqref{eq:dCS1}
\begin{align}\label{eq:dCS2}
|\alpha \rangle_d = \left[ 1+ \left( \dfrac{|\alpha|^4}{24} - \dfrac{|\alpha|^2}	{6}\alpha  a^+ + \dfrac{1}{8} \alpha^2 a^{+2} \right)\right]|\alpha\rangle \ ,
\end{align}
where $|\alpha\rangle$ is the non-deformed CS.\\
%

Furthermore, using the relations
\begin{align}
e^{\bar\beta a}a^+ &= (a^+ + \bar\beta)e^{\bar\beta a} \ ,  \hspace{0.5cm} \beta \in \mathbb{C} \ \ ,\\
a\ e^{\beta a^+} &= e^{\beta a^+} (a+\beta) \ ,
\end{align}
we can derive the following overlaps :
\begin{align}\label{eq:overlap}
_d\langle \beta|\alpha \rangle _d 
			=& \left[1 + \varepsilon \left( \dfrac{|\alpha|^4 + |\beta|^4}{24} - \dfrac{|\alpha|^2 + |\beta|^2}{6}\bar{\beta}\alpha + \frac{1}{4} \bar{\beta}^2\alpha^2 \right) \right] \nonumber \\
			& \times \langle\beta|\alpha \rangle \ ,
\end{align}
\begin{equation}
\langle \beta|\alpha \rangle _d = \left[ 1 + \varepsilon \left( \dfrac{|\alpha|^4 }{24} - \dfrac{|\alpha|^2 }{6}\bar{\beta}\alpha + \frac{1}{8} \bar{\beta}^2\alpha^2 \right) \right] \langle\beta|\alpha \rangle \ ,
\end{equation}
\begin{equation}
{}_d\langle \beta|\alpha \rangle = \left[1 + \varepsilon \left( \dfrac{|\beta|^4 }{24} - \dfrac{|\beta|^2 }{6}\bar{\beta}\alpha + \frac{1}{8} \bar{\beta}^2\alpha^2 \right) \right] \langle\beta|\alpha \rangle \ ,
\end{equation}
where $\langle\beta|\alpha \rangle = \exp[\frac{1}{2}(2\alpha\beta^* - |\alpha|^2 - |\beta|^2)]$.\\

Note that from Eq.\eqref{eq:overlap} one has
\begin{equation}
_d\langle \alpha | \alpha \rangle_d = \langle \alpha | \alpha \rangle = 1 \ .
\end{equation}
that is our DCS are naturally normalized without any need to an additional normalization constant. 
This is a specific feature compared to many other constructions of  DCS \cite{Manko1997,Recamier2008,Manko2000,Roy2000a}.
Also, Eq.\eqref{eq:overlap} shows that our $q$-deformed coherent state are non-orthogonal as in the non-deformed case.\\

It is very important to mention, as it is pointed out in Ref.\cite{Boucerredj2005}, that the DCS constructed from deformed displacement  operator and those constructed as annihilation operators eigenstates  are not the same even though the two approaches are equivalent in the non deformed formalism.\\  

The analogy between our DCS and NCS reveals the physical interpretation of the algebra deformation introduced in the previous section. Indeed, we can understand the small correction introduced to the commutation relation \eqref{eq:d-WH2}, implying a symmetry deformation, as a decoherence effect due to the system environment and acting on its physical properties. In the next section, we will study the effect of such a deformation on some properties of a system of entangled coherent states.

\section{Entanglement of deformed coherent states}\label{sec:EDCS}

The relevance of our DCS constructed from a deformed symmetry (deformed Weyl-Heisenberg algebra) resides in their interpretation as NCS which are originated from a certain kind of nonlinear potential describing an external environment action like. 
In other words, the weak $q-$deformation of the original symmetry can be explained as a small perturbation acting on the system characterized by an order parameter $\varepsilon$ and leading to a sort of  decoherence phenomenon due to the environment 
affecting the entanglement between physical states as it will be emphasized in what fellows.\\

Since the seminal works of Tombesi and Mecozzi \cite{Mecozzi1987,Tombesi1987} in which entangled coherent states (ECS) have been first studied as entities of physical interest in their own right, ECS have continued to be of a large interest in quantum information processing like quantum teleportation, superdense coding, quantum key distribution and telecloning \cite{Berrada2013,Bennett1993,Bennett1992,Ekert1991,Murao1999}.
Entangled nonorthogonal states have attracted much attention in quantum cryptography \cite{Fuchs1997}. Bosonic, SU(2) and SU(1,1) ECS are typical examples of such states \cite{Sanders1992,vanEnk2001, Hirota2001,Jeong2001,Wang2000}. \\


In what fellows we study a bipartite entanglement of DCS quantified using the entanglement concurrence. 
Let us first take a deformed state
\begin{align}\label{eq:psi2}
\mid \psi \rangle_d = \mu|\alpha \rangle_d  \otimes |\beta\rangle_d + \nu| \gamma\rangle_d \otimes |\delta \rangle_d \ ,
\end{align}
where $|\alpha \rangle_d, | \gamma\rangle_d$ (resp. $| \beta \rangle_d$ and $|\delta \rangle_d$) are {normalized} deformed coherent states of system 1 (resp. system 2) with complex coefficients  $\mu$ and $\nu$.
Following Ref. \cite{Wang2002} we define a deformed orthogonal basis $\{||0\rangle_d, ||1\rangle_d\}$ as:
\begin{align}
||0\rangle &= | \alpha\rangle_d, \hspace{0.3cm} ||1\rangle = (\mid\gamma\rangle_d -p_1|\alpha\rangle_d) / N_1 \hspace{0.3cm} \text{ for system 1},\\
||0\rangle &= | \beta\rangle_d, \hspace{0.3cm} ||1\rangle = (|\delta\rangle_d -p_2|\beta\rangle_d) /N_2 \hspace{0.3cm} \text{ for system 2},
\end{align}
where
\begin{align}
p_1 = {}_d\langle \alpha|\gamma\rangle_d \ , \hspace{1.5cm} N_1 = \sqrt{1- | p_1|} \ ,\\
p_2 = {}_d\langle \beta|\delta\rangle_d  \ ,\hspace{1.5cm} N_2 = \sqrt{1- | p_2|} \ .
\end{align}
After deriving the reduced density matrix $\rho_{1(2)}$ (resp. $\rho_{2(1)}$) for system 1 (resp. system 2) \cite{Mann1995} 
the concurrence $\mathcal{C}$ takes the well known form  \cite{Hill1997,Wootters1998}:
\begin{align}\label{eq:C1}
\mathcal{C} = \dfrac{2|\mu| |\nu| \sqrt{1-| p_1|^2} \sqrt{1-| p_2|^2}}{|\mu|^2+ |\nu|^2 + \mu\nu^{*}p_1^{*}p_2 + \mu^{*}\nu p_1 p_2^{*}} \ .
\end{align}

For general bipartite non-orthogonal pure  states 
the necessary and sufficient conditions for a maximal entanglement, i.e. $\mathcal{C}=1$, have been found and discussed in details in Refs. \cite{Fu2001,Wang2001,Wang2002}.  For the case of our interest one can show that the maximal entanglement conditions hold too. That is, for the deformed state $|\psi\rangle_d$ to be maximally entangled state the following conditions must be satisfied: 
\begin{equation}\label{cond1}
\mu = \nu e^{i\theta} \hspace{0.3cm} (\theta \in \mathbb{R}) \ ,
\end{equation}
and
\begin{equation}\label{cond2}
|\alpha|^2 + |\gamma|^2 - 2\alpha^*\gamma = |\beta|^2 + |\delta|^2 - 2\beta^*\delta - 2i(\theta+\pi) \ .
\end{equation}
This highlights the fact that 
%
{if non-deformed  CS are maximally entangled states they remain  maximally entangled states in the $q-$deformed case. That is maximally entangled coherent states are robust against algebra deformation.}\\

Furthermore, in comparison with the non-deformed case one can construct more maximally entangled deformed coherent states. As an example one has the states
\begin{align}
& |\alpha\rangle_d  \otimes |-\alpha\rangle_d - |-\alpha\rangle_d 		\otimes |-3\alpha\rangle_d \ ,\\
& |\alpha\rangle_d  \otimes |-\alpha\rangle_d - |-i\alpha				\rangle_d \otimes |i\alpha\rangle_d \ ,\\
& |\varepsilon \alpha\rangle_d \otimes |- \alpha\rangle_d - |-\alpha\rangle_d \otimes |\varepsilon\alpha\rangle_d \ ,
 \end{align}
and for $\alpha, z, z' \in \mathbb{R}$, one can find 
\begin{equation}
|\alpha\rangle_d \otimes |-\alpha + z\varepsilon\rangle_d - |-\alpha+ (z'-z) \varepsilon\rangle_d \otimes |-3\alpha+z'\varepsilon \rangle_d \ ,
\end{equation}
etc \ldots 

\section{Numerical study}\label{sec:num}

\begin{figure}[t!]
\centerline{
\begin{minipage}[d!]{6cm}
\includegraphics[scale=0.3]{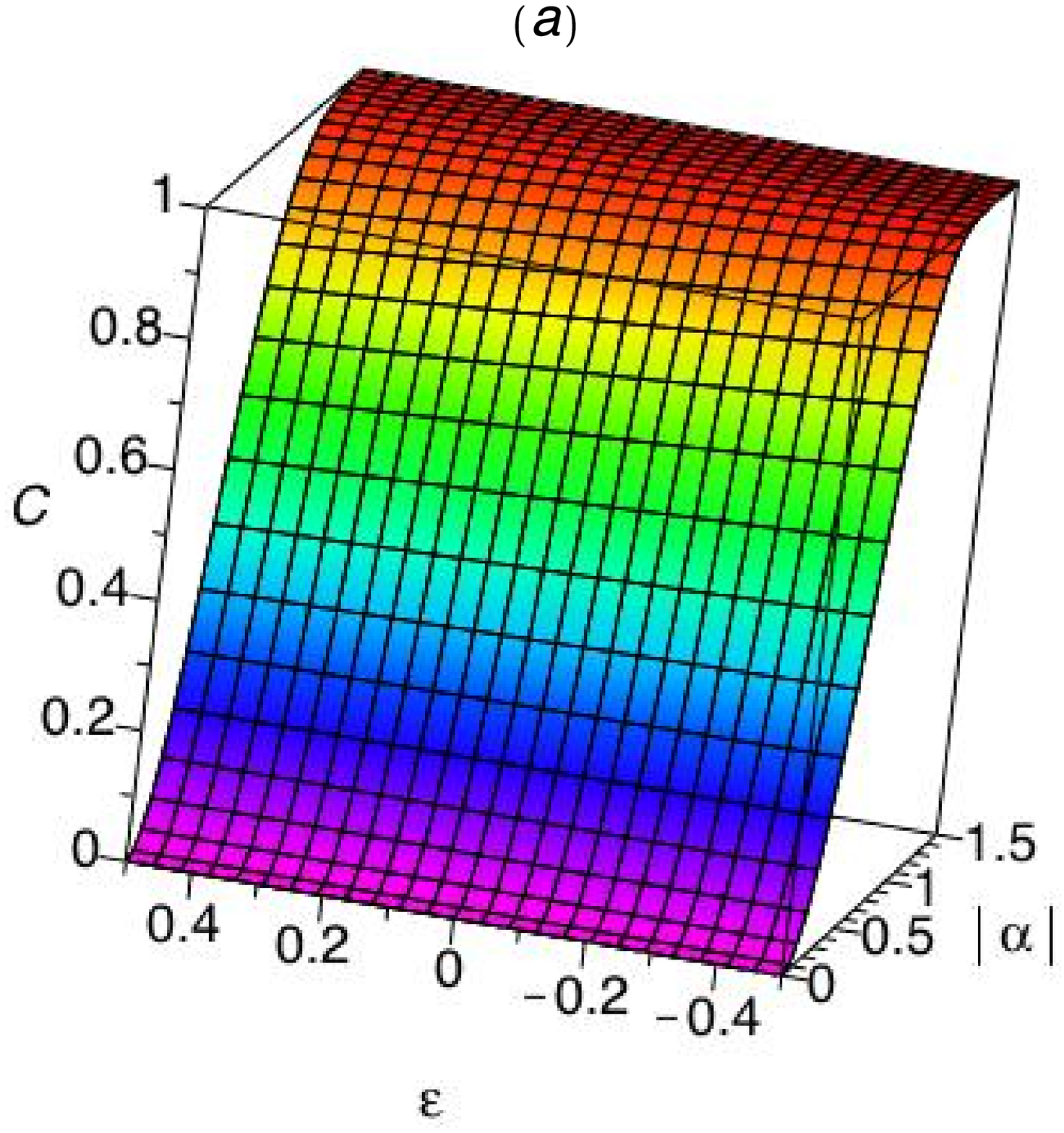}
\end{minipage}
\begin{minipage}[d!]{6cm}
\includegraphics[scale=0.3]{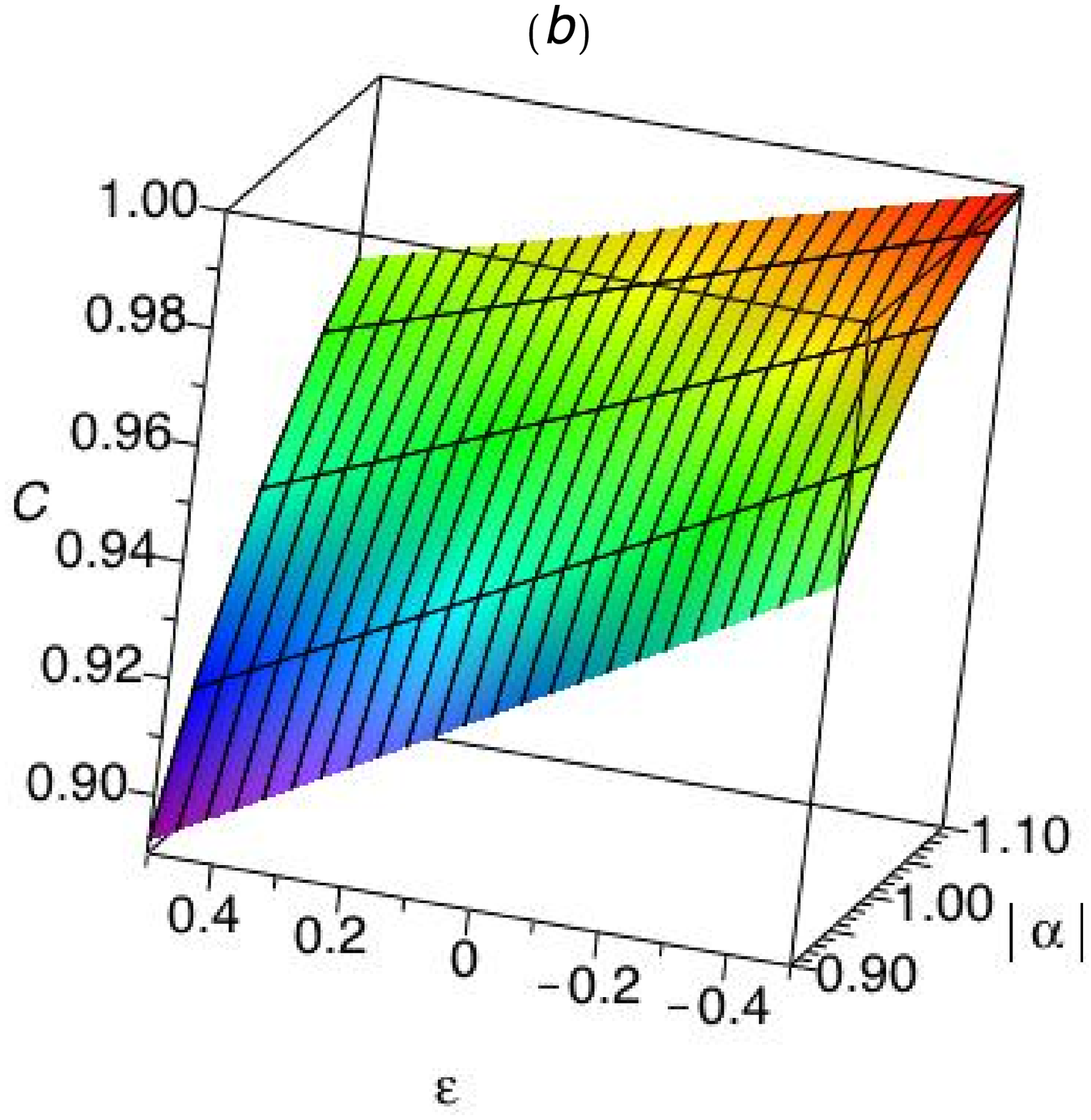}
\end{minipage}
}
\caption{(a) Concurrence of the symmetric state |$\psi_2\rangle_d(\theta=0)$ given by Eq.\eqref{eq:state2} as function of the coherence parameter absolute value $|\alpha|$ for some negative and positive values of $\varepsilon$.  (b) Zoom on the interval $|\alpha| = [0.9,1.1]$. 
}
\label{f:c2}
\end{figure}

For deeper understanding of the entanglement between two entangled deformed coherent states (EDCS) we consider states of the form
\begin{align}\label{eq:state1}
|\psi_1\rangle_d = |\alpha\rangle_d \otimes |\beta\rangle_d + e^{i\theta}| \beta\rangle_d \otimes |\alpha\rangle_d \ ,
\end{align}
for which the concurrence is given by
\begin{align}\label{eq:C2}
\mathcal{C} = \dfrac{1 - | _d\langle\alpha|\beta\rangle_d|^2}{1 + \cos \theta | _d\langle\alpha|\beta\rangle_d|^2} \ .
\end{align} 
Then, the state $|\psi_1\rangle_d$ satisfies one ebit of entanglement, i.e. $\mathcal{C}=1$, if one of the following conditions is satisfied:
\begin{enumerate}
\item[a)]
The state $|\psi_1\rangle_d$ is an antisymmetric state, i.e. $\theta = \pi$, then $\mathcal{C}=1$ independently from the parameters $\alpha$ and $\beta$ involved.
\item[b)]
The deformed coherent states $|\alpha\rangle_d$ and $|\beta\rangle_d$ are almost orthogonal, i.e. $_d\langle\alpha|\beta\rangle_d \sim 0$ (for large $\alpha, \beta$),  then $\mathcal{C}=1$ independently from the phase $\theta$.
\end{enumerate}


To be more explicit let us take the special case
\begin{align}\label{eq:state2}
|\psi_2\rangle_d = |\alpha\rangle_d \otimes |-\alpha\rangle_d + e^{i\theta}| -\alpha\rangle_d \otimes |\alpha\rangle_d \ .
\end{align}
For such an entangled state, in the weak deformation approximation limit, the concurrence given by Eq.\eqref{eq:C2} takes the form
\begin{align}\label{eq:C3}
\mathcal{C} = \dfrac{1 - \left(1+ \frac{4}{3}|\alpha|^4\varepsilon\right)e^{-4|\alpha|^2}}{1 + \cos\theta\left(1+ \frac{4}{3}|\alpha|^4\varepsilon\right)e^{-4|\alpha|^2}}.
\end{align}
Numerical calculations performed using the above concurrence of the bipartite entangled deformed coherent states $|\psi_2\rangle_d$ reveal that: 
\begin{enumerate}[(i)]
\item
{\it Maximally ECS $(\mathcal{C}=1)$ are robust against algebra deformation, confirming the analytical calculations.}

\begin{figure}[t!]
\centerline{
\includegraphics[scale=0.3]{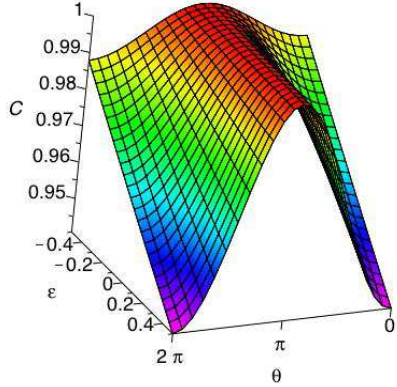}
}
\caption{Concurrence of the state $|\psi_2\rangle_d$  given by Eq.\eqref{eq:state2} as function of the phase $\theta$ for small negative and positive values of deformation parameter $\varepsilon$ corresponding to an intermediate coherence parameter $|\alpha| =1$. 
}
\label{f:c3}
\end{figure}

As a result, if non-deformed ECS ($\varepsilon=0$)
 are maximally entangled, either for the antisymmetric states i.e. $\theta = \pi$ ($\forall~ \alpha$) or for the case $\alpha \gg 1$ ($\forall~ \theta$), they remain maximally entangled independently from the deformation parameter $\varepsilon$ (see Fig.\ref{f:c2} and Fig.\ref{f:c3} ).
\item
{\it There exists an intermediate regime $(|\alpha| \sim 1)$ where $\mathcal{C}$ increases with the absolute value of the  coherence parameter $|\alpha|$ and is sensitive to the symmetry deformation parameter $\varepsilon$.}

As Fig.\ref{f:c2}a shows, the concurrence increases with increasing  $|\alpha|$ approaching a maximally entanglement with $\mathcal{C}=1$ for a sufficiently large $|\alpha|$ (depending on the value of $\varepsilon$). 
Notice that for a given value of $|\alpha|$ before the concurrence saturation, the entanglement between deformed coherent states is very sensitive to the algebra deformation. Indeed, when $\varepsilon$ increases in the interval $[-0.4,0.4]$, we notice e.g. a 4.7\% decrease of the concurrence given by Eq.\eqref{eq:C3} for $|\alpha| = 1$, and a $6.3\%$ decrease  for $|\alpha| = 0.9$  (see Fig.\ref{f:c2}b). 
\item
{\it As a function of the phase parameter $\theta$ and the deformation parameter $\varepsilon$, the concurrence which has a maximum ($\mathcal{C}=1$) at $\theta=\pi$ independently from the value of $\varepsilon$ (see Fig.\ref{f:c3}), decreases faster as  $\varepsilon$ get larger when $\theta \neq \pi$.}
%

Indeed, for the case $\theta=0$ exhibited in Fig.\ref{f:c2} we notice that if $\varepsilon$ lies in the interval $[-0.4,0.4]$, the concurrence significantly decreases by  $\sim 6.3\%$ for $|\alpha| = 0.9$, $ \sim 4.7\% $ for $|\alpha| = 1$ and $ \sim 3\% $ for $|\alpha|=1.1$. Similarly, we observe in Fig.\ref{f:c3} a reduction of $\sim 4.7\%$ when $\varepsilon \in [-0.4,0.4]$ and $|\alpha| = 1$ at $\theta = 2\pi$.
\end{enumerate} 

 The decrease of the concurrence as a function of the deformation parameter observed above can be interpreted as an effect of the environment decoherence since $\varepsilon$ plays the role of a perturbation and dissipation  parameter as it is pointed out in Sec.\ref{sec:EDCS}.\\
 
\begin{figure}[t!]
\centerline{
\includegraphics[scale=0.3]{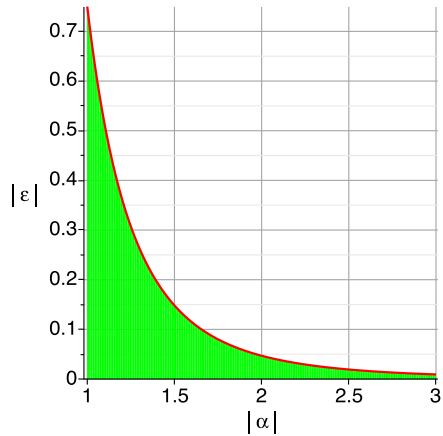}
}
\caption{Allowed values (green area) of the coherence parameter $\alpha$  and deformation parameter $\varepsilon$.}
\label{f:alp_eps}
\end{figure}

It is worth to mention that in order to have a convergent perturbation series with respect to the algebra deformation parameter $\varepsilon$ and then reliable conclusions one has from Eq.\eqref{eq:C3}  the constraint $\frac{4}{3}|\alpha^4\varepsilon| \ll 1$. Fig.\ref{f:alp_eps} displays the allowed regime for the coherence parameter $\alpha$  and the deformation parameter $\varepsilon$.

\section{Conclusion}
Through out this paper we have constructed  normalized deformed coherent states in the weak deformation approximation and shown their relationship to their nonlinear  analogues. 
We have studied the effect of such a deformation on the coherent states entanglement. 
It turns out that in comparison with the non-deformed case, the number of maximally bipartite entangled deformed coherent states is larger as new maximally entangled states can be found. 
In this paper we have quantified the entanglement using the concurrence $\mathcal{C} $ showing that maximally entangled coherent states are robust against algebra $q-$deformation. 
Moreover, numerical results reveal an important effect of the algebra $q-$deformation on the entanglement. Indeed, for bipartite entangled deformed coherent states, except the case of the antisymmetric state, the concurrence is shown to be a decreasing function of the deformation parameter for a given values of the phase and the coherence parameters. 
This can be  interpreted as a reliable argument that the algebra deformation parameter $\varepsilon$ can play the role of a decoherence order parameter representing the effect of the environment on the entangled coherent states system. Further more, preliminary results show that bipartite entanglement of three modes entangled coherent states is also affected by algebra deformation even when those later are maximally entangled states   (under investigation \cite{Rouabah2020}).

\section*{Acknowledgments}
We are grateful to the Algerian ministry of higher education and research for the financial support. M.T.R. is partially supported by an Averroes exchange program.

\begin{appendices}
\section{Derivation of deformed coherent states given by Eq.\eqref{eq:dCS2} } \label{apdx:DCS}
We define a {\it normalized} deformed coherent states by the application of the deformed displacement operator $D_d(\alpha)$ upon the vacuum state
\begin{align}\label{eq:apdx-alpha1}
| \alpha \rangle_d &=  D_d(\alpha)| 0\rangle_d \ .
\end{align}
Then, we introduce an additional term $\exp\left(\bar\alpha b\right)$  
using the fact that $e^{\bar\alpha b}|0\rangle_d = |0\rangle.$ Eq.\eqref{eq:apdx-alpha1} can be rewritten as 
\begin{align}\label{eq:apdx-alpha2}
|\alpha\rangle_d  
		& = \exp\left(\alpha b^+ - \bar{\alpha}b\right) \exp\left(\bar\alpha b\right)|0\rangle_d \ .
\end{align}

We point out that  the deformed and non-deformed vacuum states $|0\rangle_d$ and $|0\rangle$ are equivalent (see Ref.\cite{Vinod1994}). This can be checked easily in our case. Let us consider the state $|0\rangle_d = \kappa\left(|0\rangle + \varepsilon | \xi \rangle \right)$ where $|\xi\rangle$ is an arbitrary ket and $\kappa$ is a normalization constant. Applying the deformed bosonic annihilation operator on the deformed bosonic vacuum states gives
\begin{align}
b |0\rangle_d &= \kappa\left(a + \frac{\varepsilon}{4} a^+a^2 \right)\left(|0\rangle + \varepsilon | \xi \rangle\right) \ ,\nonumber\\
				& =0 \ .
\end{align}
This leads to $| 0 \rangle_d = \kappa(1+\varepsilon)| 0\rangle$. Imposing the normalization condition ${}_d\langle0|0\rangle_d = 1$, we deduce that
\begin{equation}
| 0\rangle_d = | 0 \rangle \ .
\end{equation}

To derive the expression of the deformed displacement operator coherent states given by Eq.\eqref{eq:dCS2}, we need to use the BCH formula  \cite{Riley2006}:
\begin{align}
\log \left( e^Xe^Y \right) &=  X+Y+\dfrac{1}{2!}[X,Y] + \dfrac{1}	{3!}[X,[X,Y]]/2 \nonumber \\
		&- \dfrac{1}{3!}[Y,[X,Y]]/2 - \dfrac{1}{4!}[Y,[X,[X,Y]]] \nonumber \\
		&+ \dfrac{1}{5!}\left([[[[X,Y],Y],]Y],X] +
		[[[[Y,X],X],X],Y]\right) \nonumber \\
		&+\dfrac{1}{360}\left([[[[X,Y],X],]Y],X]+ [
		[[[Y,X],Y],X],Y]\right) \nonumber \\ 
		&- \dfrac{1}{720}\left([[[[X,Y],Y],Y],Y]+ 
		[[[[Y,X],X],X],X]\right) \nonumber \\
		& + \cdots  
\end{align}
where
\begin{align}
X &= \alpha b^+ - \bar{\alpha}b\ , \hspace*{1cm}  Y = \bar\alpha b\ .
\end{align} 
Straightforward but tedious calculations  give
\begin{align}
[X,Y] 	  &=-|\alpha|^2 (1+\varepsilon b^+b) + \mathcal{O}(\varepsilon^2) \ , \\
[X,[X,Y]] &= ~|\alpha|^2 \varepsilon~ (\alpha b^+ + \bar\alpha b) + \mathcal{O}(\varepsilon^2) \ , \\
[Y,[X,Y]]  &= -|\alpha|^2 \bar{\alpha} \varepsilon~ b + 							\mathcal{O}(\varepsilon^2) + \mathcal{O}(\varepsilon^2) \ , \\
[Y,[X,[X,Y]]]  & = | \alpha|^4 \varepsilon  + \mathcal{O}(\varepsilon^2) \ ,
\end{align}
where we used the identity [A,BC] = [A,B]C + B[A,C] and neglected all terms of second and higher order in $\varepsilon$ ($\varepsilon \ll 1$).
Notice that the last commutator above corresponding to the $6^{th}$ term in BCH formula is a “c-number". That is, all next commutators vanish.\\

Using the commutation relations above, the deformed coherent states of  Eq.\eqref{eq:apdx-alpha2} become 
\begin{align}\label{eq:qDCS3}
 |\alpha \rangle_d 
 	&=  e^{-\frac{|\alpha|^2}{2}} \ e^{ - \varepsilon\frac{|\alpha|^4}{24}} \nonumber\\
 	&\times \exp \bigg[ \alpha b^+ + \varepsilon \left( \frac{|\alpha|^2}{12}\alpha b^+ + \frac{|\alpha|^2}{6} \bar\alpha b - \frac{|\alpha|^2}{2}b^+b\right)\bigg]|0\rangle \ .
\end{align}
Using a Taylor series expansion of the third exponential in Eq.\eqref{eq:qDCS3} and the relations
\begin{subequations}
\begin{align}
a^na^+  &= na^{(n-1)} + a^+a^n \ , \label{eq:WH2}\\
aa^{+n} &= na^{+ (n-1)} + a^{+n}a \ , \label{eq:WH3}
\end{align}
\end{subequations}
as well as 
\begin{align}\label{eq:apdx-d-op1}
b^+ &=   a^+ + \frac{1}{4} \varepsilon  a^{\dag2}  a + \mathcal{O}(\varepsilon^2) \ ,  \\
b &=   a + \frac{1}{4} \varepsilon a^+ a^2 + \mathcal{O}(\varepsilon^2) \label{eq:apdx-d-op2}\ .
\end{align}
Direct simplifications lead to
\begin{align}
|\alpha \rangle_d =&   e^{-\frac{|\alpha|^2}{2}} \ e^{-\varepsilon\frac{|\alpha|^4}{24}}
	\sum_{n=0}^\infty \dfrac{\left[A + \varepsilon B \right]^n}{n!} |0\rangle \ ,	\\ \nonumber
			=& \sum_n\dfrac{1}{n!}A^n |0\rangle  
		+ \sum_{p=0}^{n-1}\sum_n\dfrac{1}{n!}A^{n-p-1}\varepsilon B A^p |0\rangle  +\mathcal{O}(\varepsilon^2) \ , 
\end{align}
where
$A= \alpha b^+$ and $B= \frac{|\alpha|^2}{2}b^+b + \frac{|\alpha|^2}{12}\alpha b^+ + \frac{|\alpha|^2}{6} \bar\alpha b$.\\

Now it is easy to see that
\begin{align}
\sum_n \dfrac{A^n}{n!}  | 0 \rangle_d
		&= e^{\frac{|\alpha|^2}{2}} \left( 1 + \varepsilon \frac{1}{8}\alpha^2a^{+2} \right) |\alpha\rangle \ ,\label{eq:term2qCS}
\end{align}
and
\begin{align}
\sum_{p=0}^{n-1}\sum_n\dfrac{1}{n!}A^{n-p-1}\varepsilon B A^p  &= 	\varepsilon \left( - \frac{|\alpha|^2}{6} \alpha a^+ + \frac{|\alpha|^4}{12} \right)e^{\frac{|\alpha|^2}{2}} |\alpha\rangle	\ .
		\label{eq:term3qCS}
\end{align}

Combining
\begin{align}\label{eq:term1qCS}
e^{-\varepsilon\frac{|\alpha|^4}{24}} 
		&= 1 -\varepsilon\frac{|\alpha|^4}{24} + \mathcal{O}(\varepsilon^2) \ ,
\end{align}
with results \eqref{eq:term2qCS} and \eqref{eq:term3qCS}, our deformed coherent states take the form
\begin{align}
|\alpha\rangle_d =   \left[ 1 + \varepsilon \left(\frac{|\alpha|^4}{24} - \frac{|\alpha|^2}{6} \alpha a^+ + \frac{1}{8} \alpha^2 a^{+2} \right) \right] |\alpha \rangle \ .
\end{align}

Furthermore, using the relations
\begin{align}
e^{\bar\beta a}a^+ &= (a^+ + \bar\beta)e^{\bar\beta a} \ ,\\
a\ e^{\beta a^+} &= e^{\beta a^+} (a+\beta) \ ,
\end{align}
the overlaps ${}_d\langle\beta| \alpha\rangle_d$ writes
\begin{align}
{}_d\langle\beta| \alpha\rangle_d
		=& \left[1+\varepsilon\left(\dfrac{|\bar\beta|^4+|\alpha|^4}{24} - \bar\beta\alpha\dfrac{(|\bar\beta|^2+|\alpha|^2)}{6} + \dfrac{1}{4}\bar\beta^2\alpha^2 \right)\right] \langle\beta| \alpha\rangle \ , 
\end{align}

leading to
\begin{align}
{}_d\langle\alpha|\alpha\rangle_d 
		&= 	\left[1 + \varepsilon \left( \frac{2|\alpha|^4}{24} - \frac{2|\alpha|^4}{6} + \frac{|\alpha|^4}{4} \right)\right]\langle\alpha|\alpha\rangle \ ,\nonumber\\
		&= \langle\alpha|\alpha\rangle \ ,\nonumber \\
		& =1 \ .
\end{align}
\end{appendices}

\bibliographystyle{unsrt}
\bibliography{Bibliography_EDCS_bis,BibliographyCollectScatt_bis}

\end{document}